\documentclass[aps,superscriptaddress,showpacs,twocolumn,nobibnotes]{revtex4-1}
\usepackage{graphicx,color}
\usepackage{amsmath,amsfonts}
\usepackage{amssymb}
\usepackage{hyperref}
\usepackage{ulem}

\newcommand{\ket}[1]{|{#1}\rangle}

\newcommand{\bracket}[2]{\langle#1|#2\rangle}

\newcommand{\Up}{{\uparrow}}
\newcommand{\Dn}{{\downarrow}}

\newcommand\beq            {\begin{equation}}
\newcommand\eeq           {\end{equation}}

\begin{document}

\title[Double Weyl points and Fermi arcs of topological semimetals in...]{Double Weyl points and Fermi arcs of topological semimetals in non-Abelian gauge potentials}

\author{L. Lepori}

\affiliation{Dipartimento di Fisica e Astronomia,
Universit\`a di Padova, Via Marzolo 8, 35131 Padova, Italy.}
\affiliation{Dipartimento di Scienze Fisiche e Chimiche, Universit\'a dell'Aquila, via Vetoio,
I-67010 Coppito-L'Aquila, Italy.}
\affiliation{INFN, Laboratori Nazionali del Gran Sasso, Via G. Acitelli, 22, I-67100 Assergi (AQ), Italy.}

\author{I. C. Fulga}
\affiliation{Department of Condensed Matter Physics, Weizmann Institute of Science, Rehovot 76100, Israel.}

\author{A. Trombettoni}
\affiliation{CNR-IOM DEMOCRITOS Simulation Center, Via Bonomea 265, I-34136 Trieste, Italy.}
\affiliation{SISSA and INFN, Sezione di Trieste,
via Bonomea 265, I-34136 Trieste, Italy.}
 
\author{M. Burrello }
\email{for correspondence: michele.burrello@mpq.mpg.de}
\affiliation{Max-Planck-Institut f\"ur Quantenoptik, Hans-Kopfermann-Str. 1, 
D-85748 Garching, Germany.}

\begin{abstract}
We study the effect of a non-Abelian $SU(2)$ gauge potential mimicking 
spin-orbit coupling on the 
topological semimetal induced by a magnetic field having $\pi$-flux 
per plaquette and acting on fermions in a $3D$ cubic lattice. 
The Abelian $\pi$-flux term gives rise to a spectrum characterized 
by Weyl points. The non-Abelian term is chosen to be gauge equivalent 
to both a $2D$ Rashba and a Dresselhaus spin-orbit coupling. As a result of the anisotropic nature of the coupling between spin and momentum and of 
the presence of a $C_4$ rotation symmetry, when the non-Abelian part is 
turned on, the Weyl points assume a quadratic dispersion along 
two directions and 
constitute double monopoles for the Berry curvature. 
We examine the main 
features of this system both analytically and numerically, focusing on its gapless surface modes, 
the so-called Fermi arcs. 
We discuss the stability of 
the system under confining hard-wall and harmonic potentials, 
relevant for the implementation in ultracold atom settings,
and the effect of rotation symmetry breaking.
\end{abstract}

\maketitle

\section{Introduction}

The study of topological phases of matter and the structure of their energy 
bands have been very active fields of research in the last decade.  
 Major results are, among others, 
the theoretical classification of such phases 
in a variety of cases and the experimental characterization of solid-state 
materials with topological properties 
\cite{hasankane,zhang11,book_intro,ludwig08,chiu15}. A remarkable feature 
of such research is the strong interplay between theoretical and experimental 
advances, with the possibility of singling out materials having the desired 
properties and then proceeding to the corresponding experimental 
implementation.

 An example of systems widely studied both in $2D$ and $3D$ is given 
by models with conical intersections in the energy spectrum, 
giving rise to a variety of interesting features and phenomena, such as 
the chiral anomaly \cite{nielsen1983}, effective Lorentz violation 
\cite{grushin2012}, emergent space-time super-symmetry \cite{lee2007}, and
fractionalization \cite{hou2007}. Additionally, these systems show a
relative robustness against transitions to superfluidity and eventual 
emergence of non-BCS superfluid states \cite{cho2012}, peculiar magneto-optical conductivity \cite{ashby2013} and the density response in the presence 
of an external magnetic field exhibiting 
a non-analytic, non-classical correction to the electronic compressibility 
and the plasmon frequency \cite{panfilov2014}.

Similar phenomena characterize gapless systems displaying topological features, a prominent example being provided by
topological semimetals. Such gapless states have band structures which mimic the main properties
of the critical points of topological insulators and superconductors. 
This is the case of the so-called Weyl semimetals 
\cite{wan11,balents11,burkov11} 
in three space dimensions ($3D$): 
gapless systems characterized by non-trivial boundary phenomena 
protected by topological properties 
of their bands.
These systems present zero-energy bulk modes with linear dispersion, 
called Weyl nodes, which are protected 
against generic local perturbations of the Hamiltonian.

The Weyl nodes have a topological origin, since they 
constitute monopoles of the Berry curvature of the band structure. 
This guarantees their stability against perturbations 
and gives rise to protected gapless surface modes, called Fermi arcs, 
that connect the projections of the Weyl nodes on the surface 
Brillouin zone \cite{wan11}. 

Beside the usual Weyl semimetals, 
characterized by a linear dispersion 
of the Weyl nodes, the introduction of extra rotational symmetries 
in the system can lead to the formation of more exotic states. 
In this case, the band-touching points can be characterized by 
quadratic or cubic dispersion relations 
along two momentum directions, and constitute monopoles of 
higher charges for the Berry connection \cite{bernevig12}.

In the case of double charge, the resulting band touching points are called 
double-Weyl nodes. These particular topological objects can appear, for example, in the spectrum of strontium silicide with strong spin-orbit coupling, as was very recently pointed out \cite{hasan15e}. Such materials combine the anisotropy in their spectrum with the  
linear behavior of the density of states as a function of the energy, and this gives rise to peculiar anisotropic transport properties \cite{chen16} and to a non-trivial anisotropic screening of Coulomb interactions \cite{lai15,jian15}. Other properties, such as their behavior in the presence of disorder, have also been considered \cite{Sbierski2016}.
From this point of view is therefore desirable to consider the properties 
of double-Weyl nodes 
in models with tunable parameters.

On the experimental level, Weyl semimetals have been recently obtained in several solid-state compounds, and their properties are presently an active focus of condensed matter physics. The main strategy for their realization has been to engineer particular materials characterized by
the breaking of spatial inversion symmetry. In this way, Weyl semimetals have been realized in compounds which include tantalum arsenide \cite{lv15a,lv15b,hasan15c} and 
niobium arsenide \cite{hasan15d}. 
Further experiments allowed to observe the appearance of gapless Fermi arcs 
on the surface of compounds such as semimetal bismuth trisodium 
\cite{hasan15b} and tantalum arsenide \cite{hasan15c} 
through angle-resolved photoemission measurements. 
Beside these solid-state setups, an alternative experimental realization of 
Weyl points, similarly based on the breaking of space-inversion symmetry, 
was obtained in photonic crystals \cite{lu15}.
We notice that so far no physical realization  of Weyl semimetals 
has been implemented by breaking 
time-reversal symmetry, 
but several theoretical proposals have been presented in this direction. 
They include the use of magnetic impurities \cite{balents11}, 
complex next-nearest-neighbour hoppings in 
cubic lattices \cite{dassarma14}, or particular patterns 
of staggered magnetic fields \cite{jiang12,delplace,law2015}. 

A different strategy to achieve a Weyl semimetal is to use,  
in a tight-binding model of fermions on a cubic lattice, a magnetic field 
such that the flux per plaquette 
is $\pi$ \cite{laughlin,Hasegawa}.  Both in $3D$ \cite{laughlin,Hasegawa} 
and in $2D$ \cite{AM} the single 
particle spectrum has conical intersections. This model 
has a natural counterpart in ultracold systems \cite{LMT}: it can be implemented with ultracold fermions in a $3D$ optical lattice with a synthetic 
magnetic field generating the $\pi$ fluxes. In this context, various experimental techniques are available to simulate magnetic fluxes \cite{dalibard11}.
Indeed, in two dimensional systems, lattices 
with $\pi$-fluxes have already been experimentally realized \cite{aidelsburger11,
Aidelsburger2013,Miyake2013,bloch14,esslinger14,bloch15}, whereas the 
corresponding implementation in $3D$ is still lacking. 
However, a recent concrete proposal to realize a $\pi$-flux per plaquette in $3D$ lattices of 
ultracold atoms shows that the required space-dependent phases can be
obtained through laser-assisted tunneling in setups which could 
be implemented with current technology \cite{ketterle14}. 
Notice for comparison that large magnetic fluxes are instead 
in general difficult to achieve in solid-state realizations.

Beyond the tunability of the experimental parameters, another important 
advantage of considering ultracold atom systems is that 
in these setups one can also study generalizations of 
the Abelian $\pi$-flux model that involve artificial non-Abelian 
gauge potentials \cite{dalibard11}, including the effects of a 
spin-momentum coupling independent on the position.
In previous works, theoretical investigations have 
shown that synthetic non-Abelian gauge potentials are 
an efficient way of implementing exotic quantum Hall systems
\cite{goldman07,goldman09a,goldman09b,burrello10,burrello11,grass13,burrello15} 
or topological phases \cite{mazza12,goldman12,lewenstein13,sun13,2Dsystem} 
in two-dimensional lattices, and that laser assisted tunneling may 
provide useful tools for the simulation of particles like massless Dirac 
fermions \cite{goldman09a}, Wilson fermions \cite{bermudez2010} or Weyl 
fermions \cite{lan11}. 

In this paper, we explore the effects of a non-Abelian $SU(2)$ 
gauge potential on the topological properties of Weyl semimetals. 
In particular we examine how 
the non-Abelian coupling modifies the Weyl fermion physics of cubic lattices 
in the presence of a $\pi$-flux in each plaquette 
\cite{LMT,ketterle14}. 
The gauge potential mimics the effect of a 
spin-orbit coupling involving only two momentum components. Similar to a Rashba or Dresselhaus spin-orbit coupling, 
it has an anisotropic nature: the motion on the $\hat{z}$ direction is decoupled from the spin. 

Our result is that this anisotropic coupling of spin and momentum leads to the formation of a novel topological semimetal featuring double Weyl points. 
The present study provides a rigorous analytical basis to the recent numerical predictions by Huang {\it et al.} \cite{hasan15e} about the formation of quadratic Weyl fermions. In Ref.~\cite{hasan15e} it was shown that the spin-orbit coupling in an inversion-breaking compound, strontium silicide, leads to the creation of these new topological objects. Here, we obtain an analogous result by adding a spin-orbit coupling, thus a specific $SU(2)$ 
gauge potential, to the Weyl semimetal induced by $\pi$ magnetic fluxes in a cubic lattice. Notably, this topological phase of matter is achieved without
breaking the spatial inversion symmetry or the time-reversal symmetry of the physical system, a fact allowed by the gauge freedom of the applied gauge potential \cite{weylgauge}.

The models that we analyze 
are well-suited to describe the behavior of ultracold fermionic gases 
loaded in optical lattices and subject to artificial non-Abelian gauge 
potentials (see, for example, \cite{lewensteinbook,goldman13rev}) which act on the inner components of the atoms. 
In turn, these cold atom systems can be realized experimentally following 
several techniques, ranging from laser-assisted tunneling to 
lattice modulations \cite{dalibard11,lin09,hauke12,aidelsburger11,Aidelsburger2013,Miyake2013,bloch14,esslinger14,bloch15}.

The proposed scheme may provide 
a  useful tool for the study of the properties of 
systems with 
double-Weyl points: it constitutes both a simple theoretical playground 
which captures their main features and is the starting point for their 
experimental realization. Therefore 
it provides a platform to study the possible modifications
of the striking phenomena associated with Weyl nodes to 
the case of systems exhibiting 
double-Weyl points, as for instance the cited chiral anomaly, 
transport effects, or the emergence of non-BCS superfluid states.

Our work is organized as follows. In Section \ref{sec:abelian} we briefly 
review the main properties of a $3D$ cubic lattice model with (Abelian) $\pi$-fluxes 
acting on a single-component Fermi gas: 
this model is known to host Weyl points \cite{laughlin,Hasegawa,AM}. 
In Section \ref{sec:u2model} we add a uniform non-Abelian gauge potential, 
mimicking the effect of spin-orbit coupling, and we 
describe the spectrum and symmetries of the system.
The latter shows two inequivalent double-Weyl points, discussed 
in Subsection \ref{sec:double_weyl}, 
as allowed by the $C_4$ rotational symmetry of the model. 
After discussing the properties of the Weyl points, 
in the Subsections \ref{fermiarcs} and \ref{fermiwaves} 
we focus on the numerical and analytical study of surface modes 
(Fermi arcs) 
in the presence of hard-wall trapping potentials. 
In Subsection \ref{asym} we investigate the effect of a 
non-Abelian potential with broken rotational symmetry.
In Subsection \ref{trap} we analyze the effect of a harmonic 
trapping potential, usually present in cold atom experiments. 
We finally summarize our findings in Section \ref{outlook}.
Further calculations are presented in Appendices \ref{app:monopole} and \ref{app:sym}, where we calculate the Berry charge for a double-Weyl point and discuss the effect of gauge symmetry breaking.

\section{Cubic lattice with $\pi$ fluxes}
\label{sec:abelian}

A well-known strategy to achieve Weyl points in a cubic lattice model is to 
introduce a gauge potential corresponding to a $\pi$-flux in each 
square plaquette.  In this Section we review the main properties of  
tight-binding Hamiltonian for fermions subjected 
to such a magnetic field. 
This model has been studied extensively, see for example 
\cite{laughlin,Hasegawa,LMT,ketterle14,weylgauge}.  We consider a single species of (polarized) fermions, while two-component 
fermionic mixtures will be considered in the next Section.

Consider the Abelian vector potential in the Hasegawa gauge \cite{Hasegawa}: 
\begin{equation} \label{api}
\vec{A}_{AB}=\pi \left(z-y,y-z,0\right),
\end{equation}
where the lattice spacing is set to $1$ for simplicity. 
The resulting magnetic field is 
$\vec B=\pi \left(1,1,1\right)$, so that each lattice plaquette is pierced by a flux $\pi$.
The tight-binding Hamiltonian of this system reads
\begin{equation}
H = - \omega \sum_{\vec{r} \, , \, \hat{j}}    \, c^{\dagger}_{\vec{r} + \hat{j}} 
e^{i \theta_{j}\left(\vec{r}\right)} 
 c_{\vec{r} } + \ \mathrm{H.c.} \, ,
\label{peierls1a}
\end{equation}
where $c^\dag_{\vec{r}}$ and $c_{\vec{r}}$ 
are fermionic creation and annihilation 
operators  at lattice site $\vec{r}$, 
$\omega$ is the hopping amplitude, $\hat{j}= \hat{x},\hat{y},\hat{z}$ are the unit vectors of the lattice, and 
\beq
\theta_{j}\left(\vec{r}\right) = \int_{\vec{r}}^{\vec{r}+\hat{j}} \vec{A}_{AB} 
\left(\vec{r}\right) \cdot  \mathrm{d} \vec{r} \, .
\label{peierls2a}
\eeq
 Using Eq.~\eqref{api} in Eq.~\eqref{peierls2a} one gets
$\theta_x = \pi\left(z-y\right)$, 
$\theta_y = \pi\left(y-z\right) + \pi/2$ and $\theta_z = 0$.
To describe this system in the translationally invariant case, 
it is useful to adopt the Hasegawa approach \cite{Hasegawa} 
by using a basis defined in momentum space: 
we divide the Hamiltonian in $2\times 2$ sub-blocks by 
considering pairs of states 
$\left\lbrace k_x,k_y,k_z \right\rbrace$ and 
$\left\lbrace k_x,k_y+\pi,k_z-\pi \right\rbrace$. 
In this basis the $2\times 2$ Hamiltonian 
can be rewritten as
\beq
\label{hmag}
H(\vec{k})=  - 2\omega  \tau_x \cos k_x +2 \omega  \tau_y \cos k_y  -2\omega\tau_z \cos k_z \, ,
\eeq
where $\tau_i$ are Pauli matrices and  $\vec{k}$ are in the magnetic 
Brillouin zone \cite{Landau}. With the gauge choice in Eq. \eqref{api} we can choose
$k_x\in\left[-\pi,\pi\right)$, $k_y\in\left[-\pi,\pi\right)$, and $k_z\in\left[0,\pi\right)$. 
The eigenstates of $\tau_x$ label the even and odd sublattices in the $\hat{y}-\hat{z}$ plane respectively. 
Only one dimension of the Brillouin zone 
 can be halved, reflecting the fact that the states come in pairs and, to 
obtain the cosine terms, we had to sum the two coupled states 
$\left\lbrace k_x,k_y,k_z \right\rbrace$ and $\left\lbrace k_x,k_y+\pi,k_z-\pi \right\rbrace$ corresponding to equivalent points in the Brillouin zone. 
The spectrum of Eq.~\eqref{hmag}, 
\beq
 \frac{E_{\pm}(\vec{k})}{2 \omega} =  \pm \sqrt{\cos^{2}{k_x}+\cos^{2}{k_y}+
\cos^{2}{k_z}} 
\label{spectr}
\eeq
shows Weyl points at 
$\frac{\pi}{2} (\pm 1, \pm 1,1)$. Notice that these Weyl points are stable also when an anisotropy is introduced in the hopping amplitudes: $H(\vec{k})=  - 2\omega_x  \tau_x \cos k_x +2 \omega_y  \tau_y \cos k_y  -2\omega_z\tau_z \cos k_z$. 
The system remains a semimetal also when local attractive 
interactions are added, up to a critical value of the interaction strength, due to the vanishing density of states \cite{sorella92}, as discussed in a variety of $2D$ and $3D$ lattices, including the $3D$ $\pi$-flux model \cite{mazzucchi13}.

We observe that the Weyl nodes can be obtained by breaking either time-reversal 
or spatial inversion symmetry,
considered in their \textit{canonical} (translational invariant) form. 
However, it has been shown that in systems with local gauge symmetries 
the breakdown of the canonical time-reversal ($\mathcal{T}$) or 
spatial inversion ($\mathcal{P}$) symmetry does not imply in general a breaking of the corresponding \textit{physical} symmetries \cite{weylgauge}. 
In this case it is possible to engineer Weyl semimetals whose dynamics is invariant under both the physical symmetries. In the following we are going to denote by $P$ and $T$ the physical symmetries.  We refer 
to \cite{weylgauge} for a discussion of the difference between 
canonical and physical time- and space-reversal symmetry for the Abelian 
$\pi$-flux potential.

\section{The anisotropic $SU(2)$ model}
\label{sec:u2model}

A  rich phenomenology is  obtained
by the introduction of a non-Abelian term in the gauge potential. 
To this purpose, we consider a two-component mixture of 
fermions on a cubic lattice subject to the vector potential 
\beq
 \label{potgen}
\vec{A}  \equiv \vec{A}_{\text{AB}} + \vec{A}_{\text{NAB}} \equiv \pi 
\left(z-y,y-z,0\right)\sigma_0 + q (\sigma_x , \sigma_y, 0) \, ,
\eeq
where $\sigma_i$ are Pauli matrices characterizing the two components of the inner degree of freedom of the atoms. The Abelian term $\vec{A}_{\text{AB}}$  is the gauge potential considered in the previous Section, given by Eq.~\eqref{api}, which now contains  the $2 \times 2$ identity matrix $\sigma_0$. 
The parameter $q$ determines the intensity of the non-Abelian term $\vec{A}_{\text{NAB}}$,
\cite{2Dsystem} which is gauge equivalent to both a $2D$ 
Rashba and a Dresselhaus spin-orbit coupling. 

The experimental realization of $3D$ optical 
lattices characterized by appropriate magnetic fluxes may be 
obtained through suitable laser-assisted tunneling in 
all the three directions (see, for example, Ref.~\cite{ketterle14}). 
Engineering the $SU(2)$ potential, however, seems to be a more 
challenging task. There exists a proposal to create such potentials 
in the absence of a lattice \cite{anderson2012}. 
Several theoretical works have focused instead on the realization of $SU(2)$ 
potentials based on laser-assisted tunneling amplitudes among different 
inner species \cite{gerbier10,lewenstein13} (see also the review \cite{goldman13rev} and 
references therein) or shaking techniques \cite{hauke12b}. Only very recently a first experimental realization of a two-dimensional $SU(2)$ gauge potential has been implemented in a Rubidium gas through an optical Raman lattice \cite{wu2016}.

In Eq.~\eqref{potgen}, the Pauli matrices $\sigma_i$ refer 
to the (pseudo)spin $1/2$ degree of freedom, 
which may represent different 
hyperfine levels in ultracold atomic setups in optical lattices (see \cite{dalibard11,goldman13rev} for reviews about artificial gauge potentials in ultracold atomic systems).
However, this pseudospin could also be interpreted as associated to a further sublattice degree of freedom if we consider a $3D$ structure composed by layers of honeycomb lattices, such as the ones realized experimentally using shaking techniques. In this case the $SU(2)$ non-Abelian potential can be obtained as the effect of next-nearest-neighbor hopping with particular phases.

The potential in Eq. \eqref{potgen} 
produces a system obeying the same discrete symmetries as in the Abelian case. 
 The system Hamiltonian can be written
in terms of generalized tunneling phases as 
\begin{equation} \label{gaugeinv}
 H=- \omega \sum_{\vec{r}\,,\,\hat{j}}  c^\dag_{\vec{r}+\hat{j},s} U^j_{s,s'}(\vec{r}) c_{\vec{r},s'} + {\rm h.c.} \, , 
\end{equation}
 here $s,s'$ label the spin states and the operators 
$U^j \in U(2)$ characterize the hopping in the $j$-direction:
\begin{align}
 U^x &= e^{iq\sigma_x+i\pi\left(z-y\right)} \,, \label{Ux}\\
 U^y &= ie^{i\pi\left(y-z\right)}e^{iq\sigma_y}\,, \label{Uy}\\
 U^z &= 1\,.
\end{align}
The canonical time-reversal symmetry $\mathcal{T}$ corresponds to a transformation 
$\vec{A} \to -\vec{A}$, which maps $U^j\to U^{j\dag}$. 
The initial and final Hamiltonians are however related by a $U(2)$ 
gauge transformation $\mathcal{U}_y c_{\vec{r},s}=\left( \sigma_{z}\right)_{s,s'} e^{i\pi y}c_{\vec{r},s'}$, such that the equation 
$\mathcal{T} H \mathcal{T}^\dag = \mathcal{U}_y^\dag H \mathcal{U}_y$ 
holds. 
Therefore, $T\equiv\mathcal{U}_y\mathcal{T}=\sigma_z \,   e^{i\pi y}\mathcal{T}$ 
defines the physical time-reversal transformation. 
The spatial inversion symmetry is instead given by the real-space transformation:
\begin{equation} \label{inversion}
x \to -x \,, \quad 
y \to  - y \,, \quad 
z \to -z \,, 
\end{equation}
and it corresponds to a unitary transformation $H(-\vec{r})= \sigma_z H(\vec{r}) \sigma_z$. Therefore ${\cal P} = P=\sigma_z$, $P$ being the physical space inversion symmetry.

Using the same Brillouin zone as in the Abelian model, 
the momentum space form of the Hamiltonian in Eq. (\ref{gaugeinv})
reads 
\begin{align}
 \frac{H(\vec{k})}{2\omega}=  - \cos q \cos{k_x} \tau_x  + \cos q \cos{k_y}\tau_y -\cos{k_z} \tau_z+ \nonumber \\
    + \sin q \sin k_x \tau_x\otimes \sigma_x 
  - \sin q \sin k_y \tau_y \otimes \sigma_y\, .
 \label{Hpianiso}
\end{align}
This Hamiltonian can be put in a more symmetric form by a momentum shift $k_x \to k_x + \pi$, equivalent to a gauge transformation. In this way a Dresselhaus spin-orbit coupling in its standard form is also obtained. Moreover, it becomes explicit that the model in Eq. (\ref{Hpianiso}) displays a 
$C_4$ symmetry corresponding to rotations of angles $\pi/2$ around the $\hat{z}$ axis (including a rotation of both $\vec\sigma$ and $\vec{\tau}$).

Diagonalizing Eq. (\ref{Hpianiso}) we obtain the following eigenvalues:
\begin{equation}
\frac{E_{\pm}(\vec{k})}{\sqrt{2} \, \omega} =  \sqrt{3  +\text{cos}  2 q \, (\text{cos} \, 2 k_x + \text{cos} \, 2 k_y) + \text{cos} \, 2 k_z \pm  \, \sqrt{b(\vec{k})}} 
\label{autov1}
\end{equation}
with 
\begin{align} 
b(\vec{k}) = \sin^2 q  \Big(4 - 2 \cos^2 q (\cos 4 k_x + \cos 4 k_y) - \nonumber \\ 
-4 \sin^2 q   \big(\cos 2 k_x +  \cos 2 k_y 
-  \cos 2 k_x  \cos 2 k_y \big) \Big) \,.
\end{align}
For $q \to 0$ the spectrum~\eqref{spectr} is retrieved.

The spectrum is composed of four sub-bands, two having positive and two negative energy, linked by a particle-hole symmetry $\mathcal{C}=\tau_y \otimes \sigma_y \mathcal{K}$, with $\mathcal{K}$ complex conjugation,
such that Eq.~\eqref{Hpianiso} transforms as:
\begin{equation}
 H ( \vec{k} )  = - \mathcal{C} \, H^* ( -\vec{k} ) \, \mathcal{C}^{-1} \, .
\end{equation}
Furthermore, as we showed before, the unitary space inversion symmetry is implemented by the operator $P=\sigma_z$ and it implies that the spectrum is symmetric under $k\to-k$.

The two-dimensional limit of the model in Eq. (\ref{gaugeinv}) 
can be addressed  considering different hopping $\omega_j$ in the 
different directions and then
simply sending to zero the hopping amplitude $\omega_z$ along 
$\hat{z}$, $\omega_z \to 0$, in all the resulting expressions 
for the Hamiltonian and their energies. 
Keeping $\omega_x = \omega_y \equiv \omega$ we do not break the $C_4$ rotational symmetry and we find that
the anisotropic model of Eq.~\eqref{Hpianiso} with $\omega_z \to 0$ 
is equivalent, via the unitary global transformation 
$U=(\tau_z+\tau_y)/\sqrt{2}$ to the model studied in \cite{2Dsystem} 
in the case of Abelian flux $\pi$. 
This mapping corresponds to the transformation from the Hasegawa 
to the Landau gauge in real space: $\vec{A} \to \vec{A} - 
\vec{\nabla} \big( (z-y) x + \frac{y^2}{2}- z y \big)$. 
In this $2D$ limit a canonical time-reversal symmetry is defined 
as ${\cal T}=\tau_x\sigma_y\mathcal{K}$. 
The system is then in the topologically non-trivial symmetry class DIII 
\cite{hasankane,ludwig08,chiu15,ludwig10} but it is a gapless system, 
corresponding to a thermal metal \cite{Fulga2012} or semimetal phase.

\subsection{Double-Weyl points}
\label{sec:double_weyl}

In the presence of the non-Abelian term $\vec{A}_{NAB}$ 
the two central bands touch at zero energy in the points 
$\vec{k}_0^{(\pm,\pm)}=\frac{\pi}{2} (\pm 1, \pm 1,1)$ of the Brillouin zone. 
Unlike the Abelian case, the energy dispersion around these Weyl nodes 
is quadratic in the $\hat{k}_x- \hat{k}_y$ plane, 
whereas it remains linear in the $\hat{k}_z$ direction. 
Therefore such nodes are
double-Weyl points. 

They can be shown to be the fusion of two simple Weyl points into a 
double monopole of the Berry flux with charge $Q=\pm 2$ 
(see Appendix \ref{app:monopole}) and they appear due to 
the $C_4$ symmetry of the Hamiltonian (\ref{Hpianiso}).
If this rotational symmetry is broken by a small perturbation, 
the double-Weyl points split into two single Weyl points.
We refer to Ref.~\cite{bernevig12} for a detailed study of the multi-Weyl 
points stabilized by point group symmetries and to
Ref.~\cite{hasan15e} for a numerical analysis of a compound of strontium silicide, 
where double-Weyl points appear due to a weak spin-orbit coupling, similar to 
the non-Abelian gauge potential we consider. A different approach to obtain multi-Weyl points has been very recently experimentally implemented in photonic crystals with time-reversal and $C_3$ rotational symmetry \cite{Chen2016}.

The zero-energy points appear only when the $\tau_z$ term in 
Eq.~\eqref{Hpianiso} vanishes, thus for $k_z= \pi/2$. 
To show this, we can decompose the Hamiltonian in Eq.  
\eqref{Hpianiso} into two anti-commuting parts: 
$H=H_{xy}\left(k_x,k_y\right) + H_z\left( k_z\right)$ 
with $H_z=-2\omega\tau_z\cos k_z$, such that 
$\left\lbrace H_{xy},H_z\right\rbrace=0$. 
This implies that the term $H_z$ couples pairs of eigenstates 
of $H_{xy}$ with opposite energies $\pm \varepsilon_{xy}$. 
Therefore, in order to obtain eigenstates of the full Hamiltonian 
at zero energy, we must consider, on one side, pairs of eigenstates 
of $H_{xy}$ with $\varepsilon_{xy}=0$ and, on the other, a vanishing operator 
$H_z$. Otherwise, a gap would open between the two states in each pair 
(see more details in Ref.~\cite{paananen14}).

To examine the system in the proximity of the double-Weyl points 
it is useful to apply a perturbative approach and expand 
the Hamiltonian in series of the momenta around these nodes. 
In this way we define an effective $2\times 2$ Hamiltonian 
which approximates the behavior of the two 
intermediate energy bands in a neighborhood of the band-touching points.

For simplicity, we focus on the point 
$\vec{k}=\left\lbrace \pi/2,\pi/2,\pi/2\right\rbrace$. 
In the basis $\ket{\tau_z\sigma_z}$, the states 
$\ket{\uparrow\downarrow},\ket{\downarrow\uparrow}$ 
have vanishing energy at the double-Weyl point, 
whereas the states $\ket{\uparrow\uparrow} \pm \ket{\downarrow\downarrow}$ 
have an energy $\pm 2 \omega \sin q$.
Once we expand all the terms in the momenta to second order, 
we may consider the Hamiltonian as a perturbation of the trivial 
Hamiltonian $H=0$ for the two states. In the basis of the two 
intermediate bands, this effective Hamiltonian reads:
\begin{equation} \label{Heff}
 H_{\rm eff} = \frac{\omega}{\sin q}\left(\tilde{k}_y^2-\tilde{k}_x^2 \right) \tau_x -2\omega \frac{\cos^2 q}{\sin q}\tilde{k}_x \tilde{k}_y \tau_y+2\omega \tilde{k}_z \tau_z, 
\end{equation}
where $\tilde{k}_i = k_i - \pi/2$.

It can be shown that $H_{\rm eff}$ defines, in proximity of the point 
$\vec{k}=\left( \pi/2,\pi/2,\pi/2\right) $, a magnetic monopole of charge 
$2$ for the Berry flux (see Appendix \ref{app:monopole}). Therefore, similarly 
to the conventional Weyl semimetals \cite{wan11}, it is expected to give 
rise to unconventional protected zero-energy surface states.

Furthermore, the effective Hamiltonian in Eq.  \eqref{Heff} is expressed in the basis of the two intermediate bands, but it is also related to the spin degrees of freedom $\sigma$ and it links the expectation value of the spin to the momentum $\tilde{k}$ around the double-Weyl points. Similarly to what happens in $1D$ \cite{mazza15} and $2D$ \cite{2Dsystem} setups with the simultaneous presence of magnetic fluxes and non-Abelian gauge potentials, the expectation value $\langle \sigma (\vec{k}) \rangle$ is an observable that, in principle, can be estimated in spin-resolved time-of-flight measurements with suitable $\pi/2$ pulses. This opens the possibility of a direct measurement of the charge of the double-Weyl points through an approximated evaluation of the spin winding numbers associated to surfaces that enclose the band touching points in momentum space.
 
\subsection{Zero-energy Fermi arcs and breaking of the gauge symmetry}
\label{fermiarcs}

\begin{figure}[tb]
\centering
 \includegraphics[width=0.8\columnwidth]{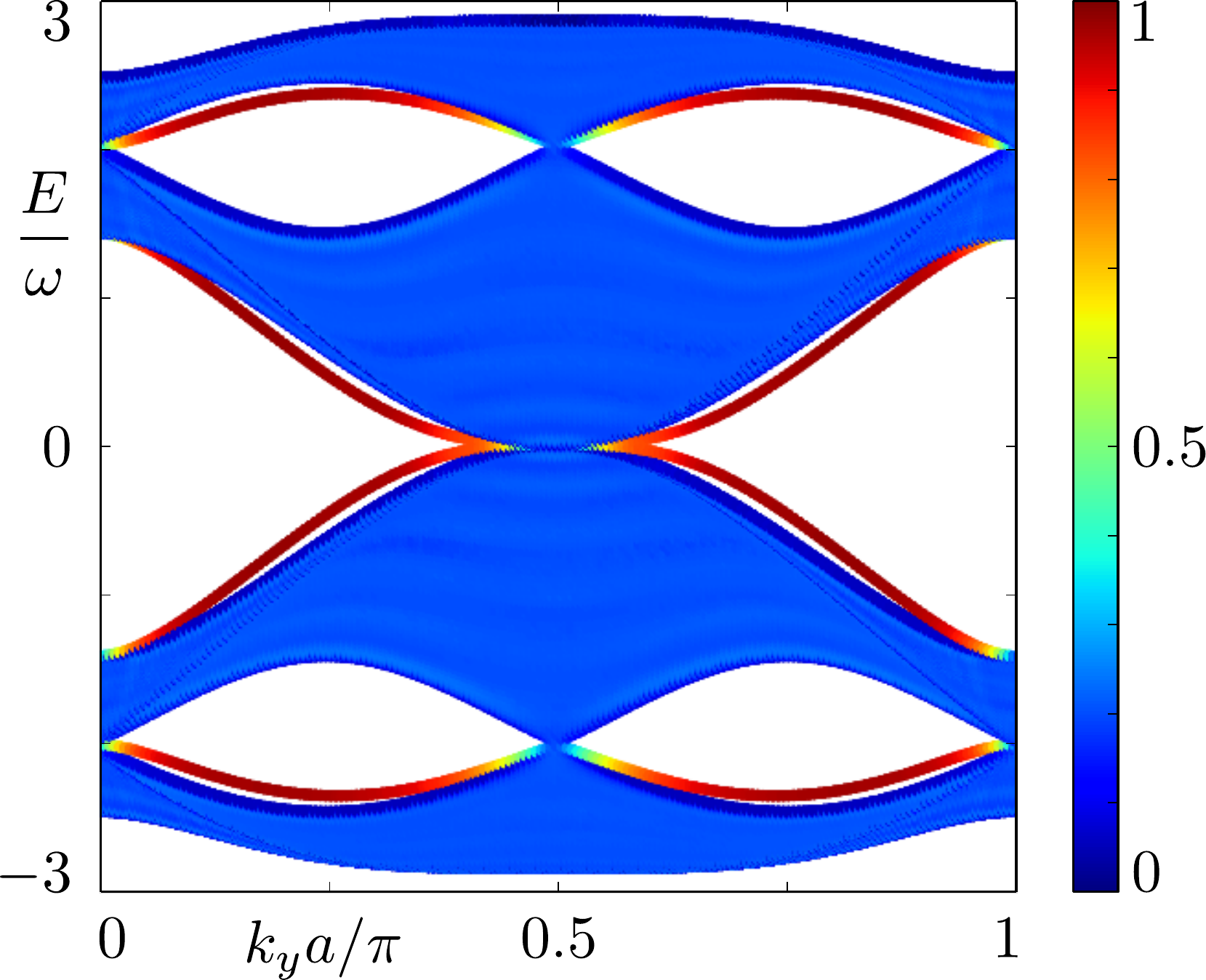}
 \caption{Energy bands in an infinite slab geometry with hard-wall boundaries along the $x$-direction, at $x=0$ and $x=120a$. We set $q=\pi/4$, and $k_z=\pi/2$. 
The color scale shows the eigenstate amplitude on the first and 
last $12$ sites from the boundaries. Double-Weyl nodes with opposite 
chirality overlap in the surface Brillouin zone, causing their Fermi 
arcs to interact and acquire a non-trivial dispersion. 
Only half of the Brillouin zone is shown, since the dispersion is 
symmetric for $k_y\to-k_y$.}\label{fig:dispersing_arcs}
\end{figure}

For both the Abelian and non-Abelian gauge-invariant models, 
the effect of a surface orthogonal to the $\hat{z}$ direction is 
trivial: no gapless surface mode can appear for a surface at $z=0$, 
as can be shown by an explicit calculation in the presence of a 
hard-wall potential.
In contrast, surfaces orthogonal to either $\hat{x}$ or $\hat{y}$ 
(the two cases are rotationally equivalent) present two localized 
surface modes which emerge from the projection of the 
double-Weyl points on the surface Brillouin zone. 
However, these states interact in pairs due to the overlap of double-Weyl 
points with opposite chirality, and no protected zero-energy Fermi arcs appear 
(see Fig.~\ref{fig:dispersing_arcs}).

The overlap of double Weyl points in the $\hat{k}_x$ and $\hat{k}_y$ 
directions is caused by two symmetries of the Hamiltonian 
in Eq.~\eqref{Hpianiso}, 
which can be expressed as translations by $\pi$ in momentum space. We get
\begin{align}
 H\left(k_x,k_y,k_z \right)&= -\tau_x H\left(k_x+\pi,k_y,k_z \right) \tau_x \,, \label{tras1}\\ 
 H\left(k_x,k_y,k_z \right)&= -\tau_y H\left(k_x,k_y+\pi,k_z \right) \tau_y \,. \label{tras2}
 \end{align}
These transformations relate Weyl points with opposite chirality through a 
translation along $\hat{k}_x$ or $\hat{k}_y$. Thus, for each double Weyl point, there exists a symmetric point which shares either the same $k_x$ or $k_y$ coordinate. Further reflection symmetries characterize the Hamiltonian \eqref{Hpianiso} and entail the same effect (see Appendix \ref{app:sym} for more details).

\begin{figure}[tb]
 \centering
 \includegraphics[width=0.8\columnwidth]{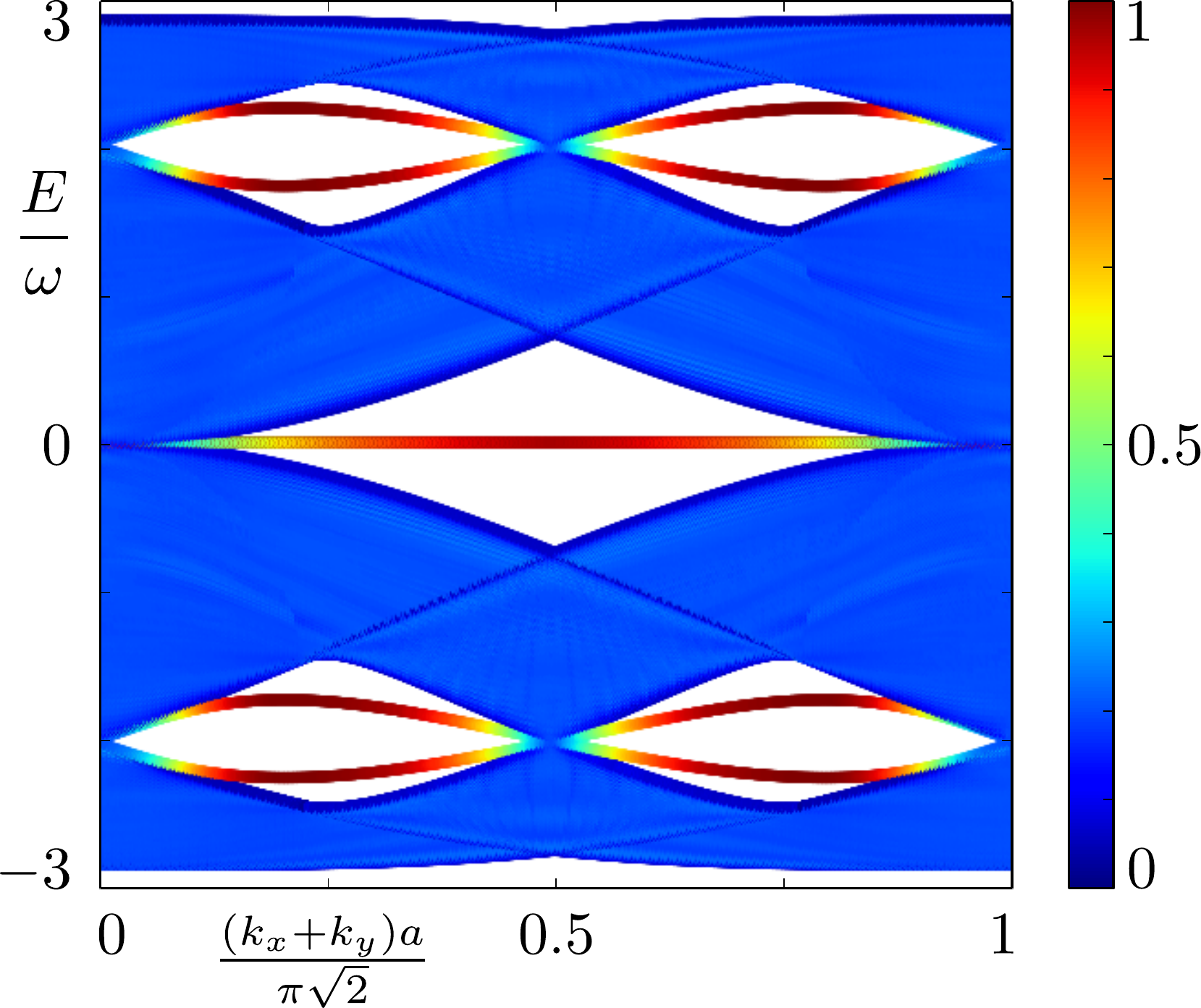}
 \caption{Bandstructure in an infinite slab geometry with hard-wall boundary conditions at $x-y=0$ and $x-y=120a$, using the same parameters 
 and the same conventions as Fig.~\ref{fig:dispersing_arcs}. 
Double-Weyl points of the same chirality overlap in the surface 
Brillouin zone, and are connected by zero-energy Fermi arcs.}
\label{fig:arcs_diag}
\end{figure}

Therefore, the system described by the potential \eqref{potgen} presents 
zero-energy Fermi arcs only on surfaces that are not orthogonal 
to the main lattice directions. 
In Fig.~\ref{fig:arcs_diag} we show the band structure of the Hamiltonian 
\eqref{gaugeinv} on an infinite slab oriented along the $\hat{x}+\hat{y}$ 
diagonal direction, with hard-wall boundaries at $x-y=0$ and $120a$. 
In this case only Weyl points with the same chirality overlap, 
leading to protected zero-energy Fermi arcs connecting their projections 
on the surface Brillouin zone and localized on the boundaries of 
the system.

In order to obtain protected zero-energy Fermi arcs also on surfaces defined 
along the main lattice directions for either $x=0$ or $y=0$, 
it is necessary to break the previous symmetries in Eqs. \eqref{tras1} and 
\eqref{tras2}.
Since they are independent on $\vec\sigma$, these symmetries are 
left untouched when varying the $SU(2)$ component of the gauge potential 
in Eq. \eqref{potgen}.

The situation becomes different if we instead consider the introduction 
of an additional term which breaks the $SU(2)$ gauge invariance and, 
therefore, the physical time-reversal symmetry. The system is then no longer 
a ${PT}$ invariant Weyl semimetal, but a usual Weyl semimetal 
with broken $T$.
An example is given by the perturbation 
$H_{xx} \equiv h_x\sigma_x \otimes \tau_x$, corresponding to a 
staggered Zeeman term
\begin{equation} \label{stagzeeman}
 H_{xx}= h_x \sigma_x \cos\left[ \pi(y-z)\right],
\end{equation}
 which may be realized by considering a weak optical superlattice of spacing $2a$ 
tuned close to appropriate anti-magic wave-lengths \cite{gerbier10}. These are special values of the wave-lengths such that the polarizability of two states of the considered atoms are opposite. Therefore, by encoding the eigenstates of $\sigma_x$ into these species, one may obtain a suitable Zeeman term in the form \eqref{stagzeeman}. A precise determination of the realization of this staggering term, however, strictly depends on the setup adopted for the implementation of the non-Abelian gauge potential. We refer to Refs.~\cite{gerbier10,lewenstein13,hauke12b,mazza12} for discussions on 
the effective engineering of these Hamiltonian terms.

\begin{figure}[tb]
 \centering
 \includegraphics[width=0.8\columnwidth]{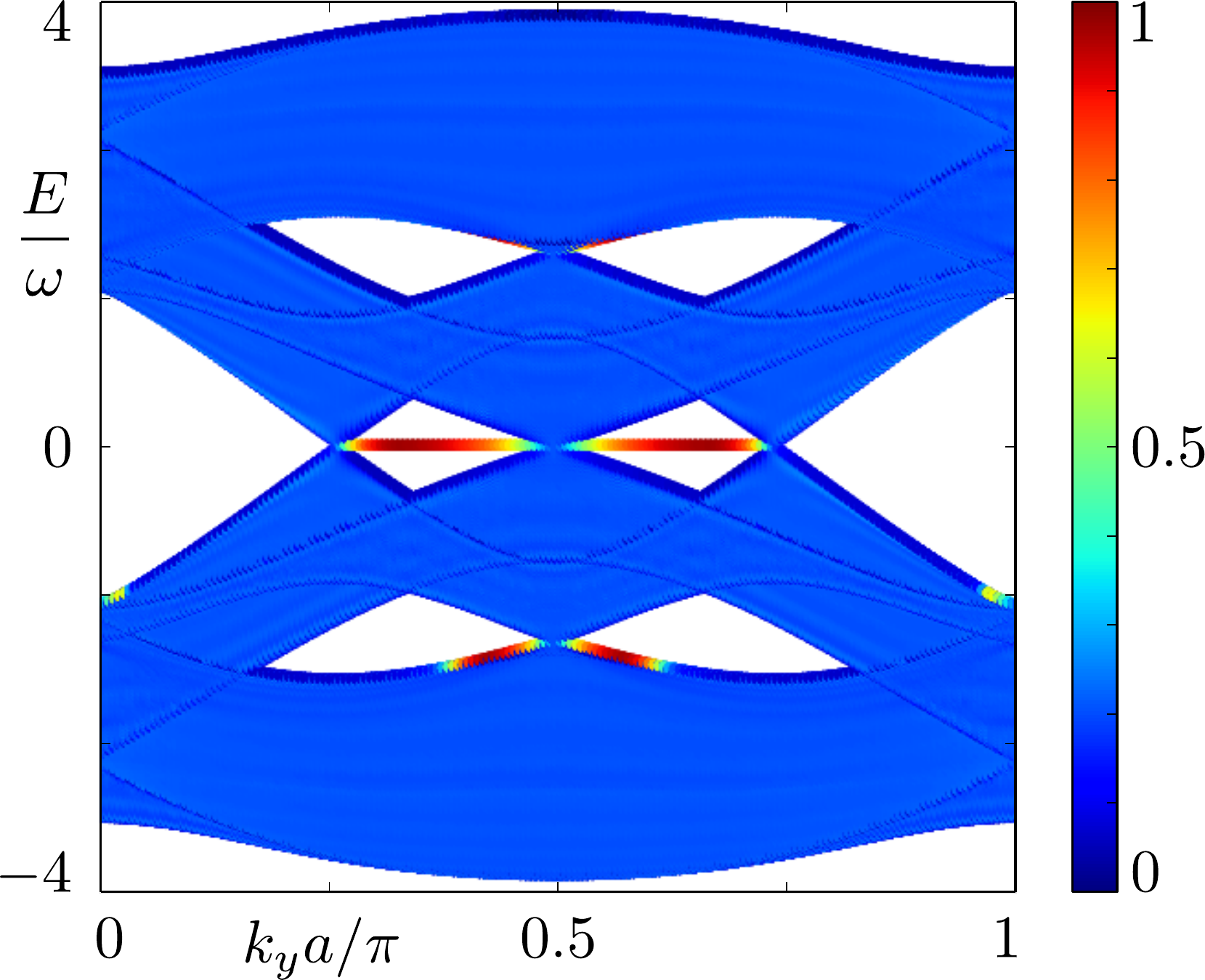}
 \caption{Same as Fig.~\ref{fig:dispersing_arcs}, but with the addition of the staggered Zeeman field $H_{xx}$ given by Eq. \eqref{stagzeeman}, with $h_x=1$.}
\label{fig:arcs_h}
\end{figure}

The perturbation Eq.~\eqref{stagzeeman} is invariant under the spatial inversion in Eq. \eqref{inversion},  
but breaks instead the physical time-reversal symmetry 
$\mathcal{U}_y{\cal T}$. The term $H_{xx}$ commutes with $H_{xy}$ 
at the double Weyl points and anti-commutes with $H_z$, thus 
avoiding the opening of an energy gap at the band-touching points. 
Additionally, $H_{xx}$ breaks both the translational symmetries along 
$k_x$ of the kind \eqref{tras1}, as well as the $k_x$-reflection symmetries 
listed in Appendix \ref{app:sym}. Therefore it leads to the formation of 
Fermi arcs on surfaces orthogonal to the $\hat{x}$ direction, as seen 
in Fig.~\ref{fig:arcs_h}. 

\begin{figure}[tb]
 \centering
 \includegraphics[width=0.95\columnwidth]{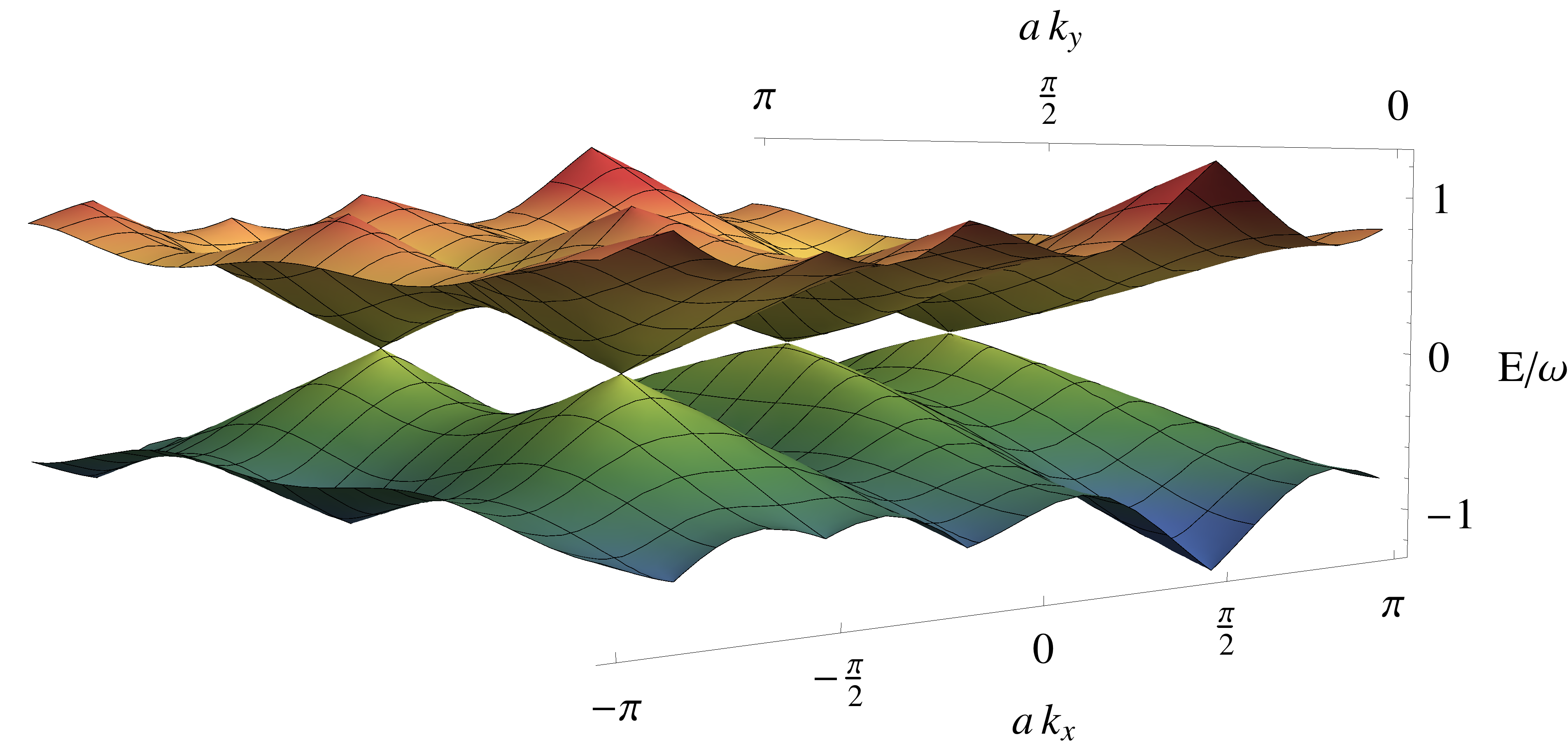}
 \caption{Energy of the two intermediate bands in the presence of the staggered Zeeman term for $k_z=\pi/2, q=\pi/4, h=1/3$. The Zeeman term splits the double-Weyl points in such a way that there are pairs of Weyl points that do not overlap once projected in the $\hat{k}_x$ direction. For clarity only half of the Brillouin zone is shown.\label{fig:split_nodes}}
\end{figure}

Finally, we notice that $H_{xx}$ breaks both the canonical particle-hole 
symmetry $\mathcal{C}$ and the $C_4$ rotational symmetry around the 
$\hat{z}$ axis. Therefore all the double Weyl points split into 
pairs of Weyl points as shown in Fig.~\ref{fig:split_nodes}.

\subsection{Fermi arcs wavefunction}
\label{fermiwaves}

In the presence of the staggered Zeeman term \eqref{stagzeeman} and of a boundary on the system orthogonal to either $\hat{x}$ or $\hat{y}$, the existence of the Fermi arcs can be shown analytically. For this purpose we start from the Hamiltonian:
\begin{multline} \label{Hfermi}
 H_{xy} + H_{xx} =  - \cos q \cos{k_x} \tau_x \otimes \sigma_0  + \cos q \cos{k_y}\tau_y\otimes\sigma_0 + \\ 
  + \left( \sin q \sin k_x +h_x\right)  \tau_x\otimes \sigma_x - \sin q \sin k_y \tau_y \otimes \sigma_y \, , 
\end{multline}
in which we fixed $\omega=1/2$ for the sake of simplicity and $k_z=\pi/2$ since it is a necessary requirement to obtain a vanishing energy, due to the condition $H_z=0$. 

First, we show that there are Weyl points located at $k_x=-\pi/2$ which have no counterpart at $k_x=\pi/2$. 
For $k_x=-\pi/2$ the Hamiltonian takes the block off-diagonal form:
\begin{equation}
 H=\begin{pmatrix}
    0 & A(k_y) \\
    A^\dag(k_y) & 0
   \end{pmatrix}\,,
\end{equation}
with
\begin{equation}
 A(k_y)=\begin{pmatrix} 
   -i\cos q \cos k_y & h + \sin q \left(\sin k_y - 1 \right)  \\
   h - \sin q \left(\sin k_y + 1 \right) & -i\cos q \cos k_y
   \end{pmatrix} \, .
\end{equation}
Its gap closes for $\det A=0$. From
\begin{equation}
 \det A(k_y)= -\cos^2 q +  \sin^2 k_y - (h-\sin q)^2 = 0
\end{equation}
we get:
\begin{equation} \label{weylcoord}
 \sin k_y = \pm \sqrt{h^2-2h\sin q + 1}\,.
\end{equation}

If $h < 2\sin q$ (we consider $0<q<\pi$) 
four Weyl points appear for $k_x=-\pi/2$. 
It is easy to verify that such Weyl points have no counterpart 
for $k_x=\pi/2$ (see Fig.~\ref{fig:split_nodes}), and therefore do not 
overlap with other band-touching points once projected on a surface 
Brillouin zone along $\hat{k}_x$.

To find a simple approximate expression for the Fermi arc wavefunctions 
appearing on the surface at $x=0$, we adopt  the approach discussed in
Refs.~\cite{paananen14,ojanen13}, starting from the Hamiltonian \eqref{Hfermi}. 
We expand it to first order in $k_x$ in a neighborhood of the Weyl points 
at $k_x=-\pi/2$ and then substitute $k_x=-i\partial_x$, since translational 
invariance along $\hat{x}$ is broken. We obtain the following equation 
for the zero-energy states $H \psi(x,k_y,k_z=\pi/2)=0$: 
\begin{align}
 \partial_x \psi(x)= \Big[ -\cos k_y\tau_z  +  i\frac{h-\sin q}{\cos q} \sigma_x  + \nonumber\\
+ \tan q \sin k_y\tau_z \sigma_y\Big]\psi(x) 
\end{align}
leading to the coupled set of differential equations
\begin{align}
 \partial_x \psi_\uparrow &= i \frac{h-\sin q-\tau_z\sin q \sin k_y}{\cos q}  \psi_\downarrow -\tau_z  \cos k_y \psi_\uparrow \label{surf1} \\
 \partial_x \psi_\downarrow &= i \frac{h-\sin q+\tau_z\sin q \sin k_y}{\cos q}\psi_\uparrow - \tau_z  \cos k_y \psi_\downarrow \label{surf2}\,.
\end{align}
From the last equation we obtain
\begin{equation} \label{fermiup}
 \psi_\Up = -i  \frac{\cos q\left( \partial_x+\tau_z\cos k_y\right) }{h-\sin q + \tau_z \sin q \sin k_y} \psi_\Dn\,.
\end{equation}
Substituting into \eqref{surf1} one has
\begin{equation}
 \left(\partial_x+\tau_z\cos k_y\right)^2 \psi_\Dn (x)
 = \frac{- \left(h-\sin q\right)^2 + \sin^2 q \sin^2 k_y}{\cos^2 q} \psi_\Dn (x)\,.
\end{equation}
Considering a solution of the kind $\psi_\Dn (x) \propto e^\alpha x$ leads to
\begin{equation} \label{alpha}
 \alpha=-\tau_z \cos k_y \pm i \sqrt{\frac{(h-\sin q)^2-\sin^2 q \sin^2 k_y}{\cos^2 q}}\,.
\end{equation}
Therefore we obtain that, for $h<\sin q$ and hard-wall boundary conditions 
at $x=0$, there are two Fermi arcs of the form
\begin{equation} \label{fermidown}
 \psi_\Dn(x) \propto \sin \left[\sqrt{\frac{\sin^2 q \sin^2 k_y - (h-\sin q)^2}{\cos^2 q}} x\right] e^{-\tau_z \cos k_y x}\,.
\end{equation}
The Fermi arc wavefunctions are zero-energy eigenstates 
for the Hamiltonian $H_{xy}$. 
By considering a value $k_z \neq \pi/2$, however, they acquire a 
dispersion given by $H_z=-2\omega\tau_z\cos k_z$. 

\begin{figure}[tb]
 \centering
 \includegraphics[width=0.8\columnwidth]{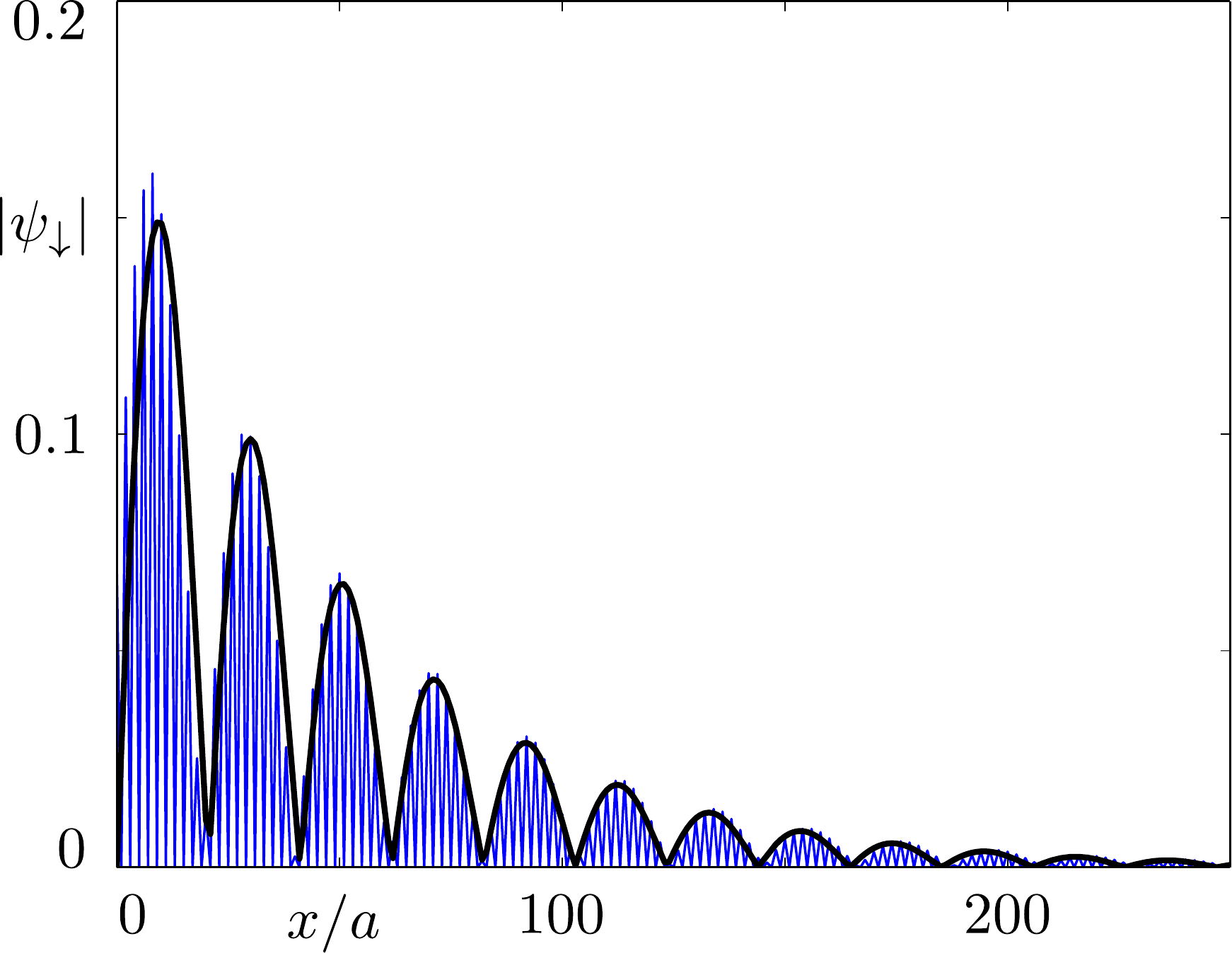}
 \caption{Comparison of numerical and analytic wavefunctions (spin down components) for the zero-energy Fermi arcs. The blue curve (thin line) is numerically obtained in an infinite slab geometry of thickness $L_x=800$, with $q=0.2$ and $h=-0.05$ at momenta $k_y=\pi/2+0.02$ and $k_z=\pi/2$ using the Hamiltonian \eqref{Hpianiso}. The thick black curve is the analytic result \eqref{fermidown}. For small values of $q$ and negative $h$ Eq.~\eqref{fermidown} provides a good approximation of the Fermi arcs wavefunctions. \label{fig:arc_wf}}
\end{figure}

The approximate solution of Eq.~\eqref{fermidown} relies on the first order 
expansion of the Hamiltonian in $k_x$ and, for example, it 
is not in exact agreement with Eq.~\eqref{weylcoord}, which defines the 
correct extension of the Fermi arcs in $k_y$. Furthermore we emphasize that the Fermi arcs appear for $h<2\sin q$, as deduced from Eq.~\eqref{weylcoord} 
and not for $h<\sin q$ as predicted by \eqref{alpha}. However, 
the approximate solution Eq.~\eqref{fermidown} allows 
to verify  analytically the existence of the zero-energy Fermi arcs 
and gives a good description of these surface states when $q \ll \pi/2$ 
and $h<0$. In this case both the exponential decay and the oscillatory 
term properly match the numerical results (see Fig.~\ref{fig:arc_wf}). 
By increasing $q$ the oscillatory term is only qualitatively captured, 
and its decay length does not match the numerical solution. 

\subsection{Effect of the breaking of rotational symmetry 
on double-Weyl points and Fermi arcs}
\label{asym}

We saw above that double-Weyl points occur for arbitrary intensity 
of the non-Abelian gauge field $\vec{A}_{\mathrm{NAB}}$ in (\ref{potgen}). 
However, a central ingredient for the emergence of double-Weyl points 
is $C_4$ rotational symmetry around the $\hat{z}$ axis which implies that 
the coefficient in front of $\sigma_x$ and $\sigma_y$ in \eqref{potgen} 
must be the same for the two components.

In real experiments, however, this assumption can be violated. 
For this reason it is important
to examine the effect of a weak perturbation breaking the rotational symmetry. Such anisotropy can be modeled by the new vector potential:
\beq
 \label{potgen2}
\vec{A} = \pi \left(z-y,y-z,0\right)\sigma_0 + (q_x \, \sigma_x , q_y \, \sigma_y, 0) \, ,
\eeq
where we consider in general $q_x \neq q_y$.

Similarly to the previous situation defined by (\ref{potgen}),
a closed expression for the four energy bands resulting from \eqref{potgen2} 
can be obtained, but the final formula for the energies is rather 
complicated and not particularly enlightening. Therefore,
we will not give it here, limiting instead the discussion to its main features.

\begin{figure}[tb]
 \centering
 \includegraphics[width=0.95\columnwidth]{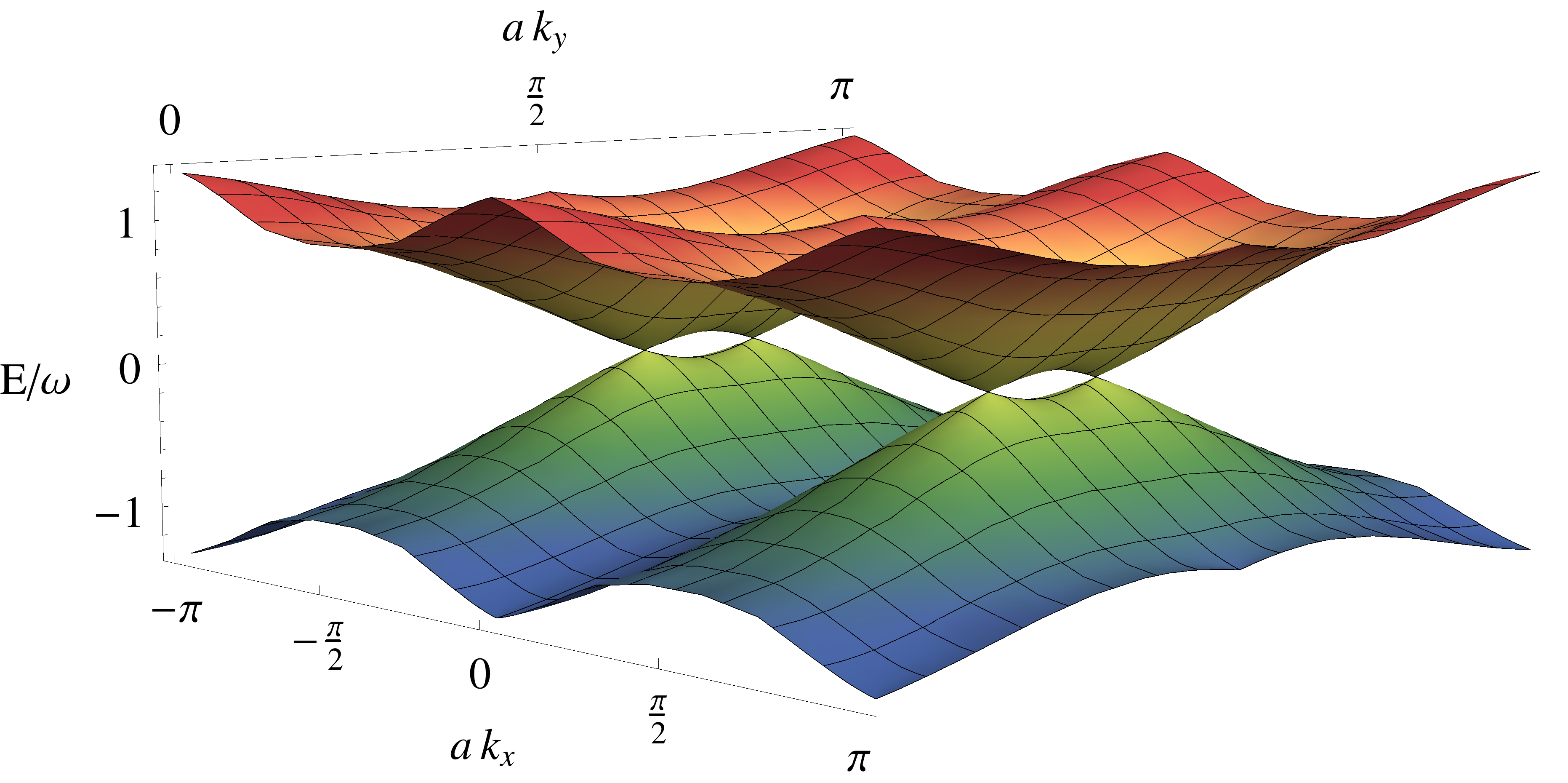}
 \caption{Splitting of a double-Weyl point in two single-Weyl points in presence of an asymmetry $q_x \neq q_y$. We set $k_z=\pi/2$, 
 $q_x=0.3$, and $q_y=0.4$. For clarity only half of the Brillouin zone is shown, corresponding to two double-Weyl points separating into two pairs of single-Weyl nodes.}
 \label{anys}
\end{figure}

The main result is that, as expected, for small anisotropies $q_x \approx q_y$, each double-Weyl point splits into two inequivalent single Weyl points 
with linear dispersion, separated along the direction corresponding to the largest between $|q_x|$ and $|q_y|$. A typical situation is shown in Fig.~\ref{anys}.

In this case, the Fermi arcs of the double-Weyl points in a diagonal slab split into 
two Fermi arcs, partially overlapping and related to the single-Weyl points. 
Since the two Weyl points are generated by the same double-Weyl node, and therefore have the same charge, there cannot be any detrimental interactions between these Fermi arcs, and the latter are pinned to zero energy.

For small perturbations given by $\delta q = |q_x-q_y|$, 
as the ones coming from imperfections in real experiments, we find that 
the distance between the Weyl points is linear in 
$\delta q$. This implies that in realistic setups such a perturbation 
could be controlled to be negligible with respect to the adopted detection 
techniques. The same considerations apply to the Fermi arcs.

We observe that a similar phenomenology would be generated by the introduction of different hopping amplitudes in the $\hat{x}$ and $\hat{y}$ directions. In the case $\omega_x \neq \omega_y$ the rotational symmetry is indeed trivially broken.
 
Concerning the experimental observations of single and double-Weyl points in ultracold atom settings, 
the main tools available with current ultracold atom technology rely on the 
possibility of detecting the shape of the energy bands and the Berry curvature 
as a function of the momenta. 

Regarding the band mapping, 
several techniques, including Bragg spectroscopy, 
have already been successfully applied in $1D$ and $2D$ systems 
(a general technique is applied, for example, in \cite{sengstock10}).
Another powerful tool is the study of Landau-Zener processes which enable to detect the gap between two bands along different paths in momentum space, based on the non-adiabatic transitions between the two bands \cite{lim12}. These measurements have been adopted to detect $2D$ Dirac cones in optical lattices with a tunable honeycomb geometry \cite{tarruell12}, and to study the phase diagram of the topological Haldane model \cite{esslinger14}. It is interesting to notice that in the previous experiments such techniques have been applied to band-touching points with a linear dispersion, corresponding to a standard Landau-Zener process as the ones expected for standard Weyl points \cite{law2015}. The study of double-Weyl points through this technique would instead require a generalization of the Landau-Zener description for a quadratic dispersion, in a way similar to the analysis of the merging transition of Dirac cones in two dimensions \cite{fuchs12}.

The measurement of the Berry curvature is another 
useful tool which can be adopted for the detection of double-Weyl points.
Berry curvatures have successfully been measured in several $2D$ systems: these observations include interferometric experiments \cite{duca15} and 
measurements of the population in different energy bands \cite{sengstock15}, 
also in the case of non-Abelian Berry connections 
in two-band systems \cite{li15}. Once extended in a $3D$ setup, 
this would allow a direct measurement of the Berry monopole for both single and multi-Weyl points.

\subsection{Zero-energy states in the presence of a harmonic trap} 
\label{trap}

The experimental setups of ultracold atomic gases loaded in optical lattices 
is usually based on either harmonic trapping potentials \cite{string} 
or the 
implementation of a hard-wall confinement for the atoms obtained through 
optical box traps \cite{gaunt13} or light-intensity masks 
\cite{corman14,corman14b}. It is therefore interesting to examine the 
behaviour of the zero-energy modes in the presence of both these potentials and to evaluate 
how they are related to the double-Weyl points.

By trapping the atoms in this way, the sharp box-shaped potentials allow us to 
define precise surfaces for the system, whereas the role of a weak 
harmonic trapping can be assimilated to a space-dependent chemical 
potential in a local density approximation regime.

Hence, in order to investigate the effect of the trapping, we assume the previous diagonal slab geometry with the addition of a shallow 1D harmonic 
potential which depends only on the distance from the surface of the slab. The system is thus infinite and translational invariant along the $\hat{z}$ and $\hat{x} + \hat{y}$ directions, but has a finite thickness along the $\hat{x} - \hat{y}$ direction. This allows us to compare the zero-energy surface states arising with and without the harmonic trap.
The single-particle Hamiltonian is
\begin{equation}
 H_{\rm trap} = H + m \, \Omega^2 (x-y)^2/2 -\mu,
\end{equation}
where $H$ is given by Eq.~\eqref{gaugeinv}.
In the  central part of the trap the harmonic potential is negligible 
and, for values of the chemical potential such that $0<\mu \lesssim  2 \omega$,
the system is in a metallic phase with the two lowest bands fully occupied 
and the third band only partially filled. In a local approximation, 
the  Fermi surface with zero energy corresponds to a set of closed lines around the double-Weyl points.

By increasing the distance $|x-y|$ from the center, 
the harmonic potential effectively reduces the local chemical potential. 
Therefore, the Fermi surface shrinks at first towards the double-Weyl points, 
until a distance such that $m \, \Omega^2 (x-y)^2 = 2\mu$ 
where only the double-Weyl points remain as zero-energy states.
Moving further towards the surfaces,  the system 
is again in a metallic phase, now with the second band partially filled. 
Here the zero-energy modes define larger and larger contours around 
the band-touching points.
Finally, on the hard-wall boundary at $|x-y|=L$, zero-energy surface modes 
appear. Their energy can be roughly estimated by:
\begin{equation}
 H_{\rm surface} \approx -2\omega\tau_z\cos k_z - \mu + m\Omega^2 L^2/2.
\end{equation}
This dispersion relation implies that the Fermi arcs are shifted by 
$\pm \delta k_z$ from their previous momentum $k_z=\pi/2$ in order to 
compensate for the trapping energy. This approximation however fails 
when properly considering the penetration of the surface modes in the bulk.

\begin{figure}[tb]
 \centering
 \includegraphics[width=0.9\columnwidth]{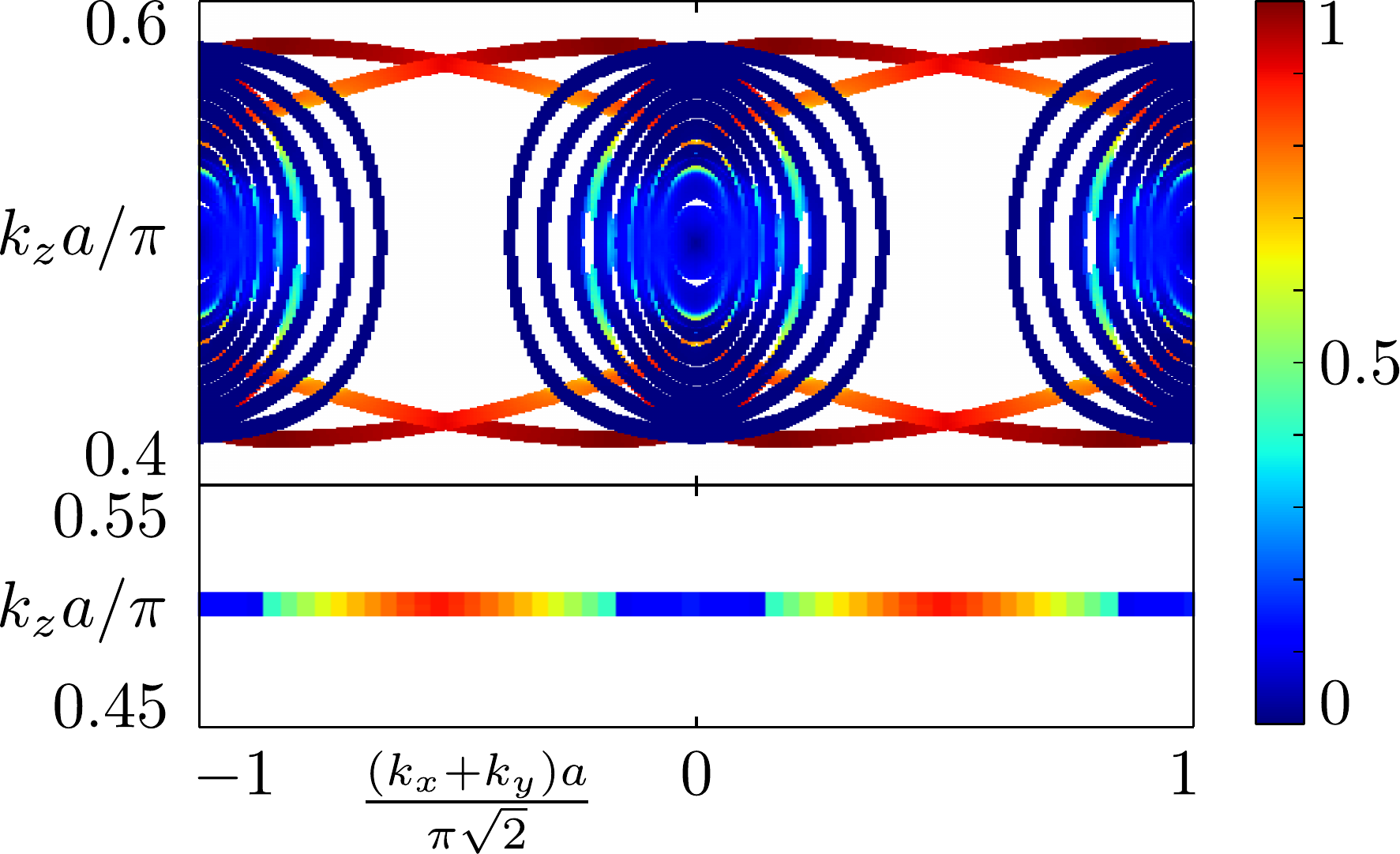}
 \caption{In the top part of the figure we plot 
in the surface Brillouin zone the states with energy $|E|<0.02\omega$ in the presence of a harmonic trapping potential. We use an infinite slab geometry 
with boundaries at $x-y=0$ and $80a$. The color scale depicts the 
eigenstate intensity on the first and last $8$ sites from the boundaries. 
The parameters $\Omega$ and $\mu$ are chosen in order to have the 
potential energy ranging from $-0.5\omega$ to $+0.5 \omega$ from 
the center of the system towards the surface. The double-Weyl points become 
extended regions of bulk states with zero energy, which are connected by 
Fermi arcs, bent by the harmonic trap. 
For comparison, the states without the harmonic trap are shown in the 
lower part of the figure.} 
\label{fermitrap}
\end{figure}

In Fig.~\ref{fermitrap} we plot the states around zero energy 
on the surface Brillouin zone for the diagonal slab geometry. 
We used a value of the trapping $\Omega$ such that the local chemical 
potential ranges from $\mu$ to $-\mu$ with $\mu<\omega$.
Our numerical results show that the gapless surface states interpolate, with 
curved shapes, between the zero-energy Fermi surfaces surrounding the projections of the double-Weyl points, 
as emerging from previous discussions \cite{wan11,ojanen13}.

This is the expected result as long as the trapping is weak enough in such 
a way that on the boundary of the system the local chemical potential 
still falls within the second band. At variance, 
if the trapping increases further such that $\mu \gtrsim \omega$, 
all the surface Brillouin zone will be covered by bulk zero-energy states, 
and the boundary of the system will be locally found in a gapped phase 
corresponding to an insulator, where only the first band is filled. 
Even larger values of $\Omega$ can then lead to the appearance of a 
new metallic phase corresponding to a partial filling of the first 
band close to the boundary.

 The zero-energy Fermi surfaces, appearing directly as states in the Brillouin zone, can be experimentally probed through a band mapping based on Bragg spectroscopy \cite{sengstock10}, analogously to the bulk states (see the end of Subsection \ref{asym}).

\section{Outlook}
\label{outlook}
A flourishing research activity about Weyl semimetals is developing in the last few years with the aim of proposing and experimentally realizing these systems and classifying the various types of Weyl nodes (see, e.g., the very recent work \cite{soluyanov15}).
In this paper we analyzed the effects of a 
non-Abelian gauge potential on a system of two-species fermions moving in 
a cubic lattice with nearest neighbor hopping only. We considered 
in particular the case of an $U(2)$ potential describing both a 
$\pi$ magnetic flux in each plaquette and a spin-orbit coupling. 
When the non-Abelian component of the potential preserves a $C_4$ rotation symmetry 
along the $\hat{z}$ axis, the system is characterized by double-Weyl-points. 
These are zero-energy band-touching points with a quadratic dispersion 
along two directions, and correspond to double monopoles for the Berry 
connection. Our findings provide a useful and simple analytical model for 
their realization and confirm the numerical results showing that the 
introduction of a suitable spin-orbit coupling leads to the formation 
of double-Weyl points.
We observe that in the non-interacting limit the double-Weyl points 
we are considering are distinctly different from a generic quadratic 
band-touching \cite{abr} due to the anisotropy of the dispersion relation, 
which enables the formation of a monopole of the Berry curvature thanks to the linear dispersion in the $\hat{z}$ direction.

The double-Weyl points lead to the formation of peculiar surface states. 
In a slab geometry oriented in a diagonal direction in the $xy$ plane, 
pairs of zero-energy Fermi arcs originate from each of these zero-energy 
bulk states. When the boundary is instead orthogonal to one of the main 
directions, $\hat{x}$ or $\hat{y}$, the surface modes interact and 
acquire a non-trivial dispersion. In this case, in order to obtain 
Fermi arcs, the breakdown of the specific symmetries is required. 
This can be achieved, for instance, by an additional staggered Zeeman term. 
However, the latter also breaks the translational and $C_4$ rotational 
symmetries, splitting the double-Weyl points in pairs of single-Weyl ones.

In the same way, any anisotropy breaking the $C_4$ rotational symmetry 
leads to a splitting of the double-Weyl points.
However, we expect that the amount of anisotropy occurring in real experiments is small, and does not lead to sensible difference from the behaviour characteristic of these points.

We finally analyzed the effects of the harmonic trapping, 
showing that it leads to the formation of topological metals 
with non-trivial Fermi surfaces, while preserving the zero-energy surface 
modes.

As an interesting avenue for future research, it would be interesting to study the effect of disorder both in the Abelian 
fluxes and in the intensity of the non-Abelian term to determine how 
robust are double-Weyl points with respect to perturbations and 
imperfections, similar to Ref.~\cite{Sbierski2016}. In the purely Abelian limit, disorder in fluxes was considered 
in \cite{weylgauge} and on the basis of these results we expect that 
the disorder in the translational invariant non-Abelian term may not play 
a crucial role and it does not add significant modifications to the effects 
of the disorder of the $\pi$-fluxes. In the non-interacting case it would be 
important to consider more general non-Abelian terms $\vec{A}_{\text{NAB}}$ 
to classify the multiple Weyl points one may obtain. It would be also 
interesting to consider the effect of 
long-range couplings, as the ones studied in \cite{gor} and associated 
to non-trivial topologies of the Fermi surface.
 
In the interacting case a first direction of future investigation would be  
to add local interaction, as in the case of alkali atom, and determine 
the differences with the superfluid states in the presence of 
quadratic band touching. Another promising future work would be to consider 
long-range interactions, possibly in the presence of short-range interactions 
(as it was recently considered in \cite{boe} for $3D$ 
fermions close to a quadratic band touching point).

{\bf Acknowledgements --} Discussions with H. Buljan are gratefully 
acknowledged. L.L. acknowledges 
financial support by the ERC-St Grant ColdSIM (No. 307688). 
I.C.F. thanks the ERC under 
the EU Seventh Framework Programme (FP7/2007-2013) / ERC Project MUNATOP, 
the US-Israel Binational Science Foundation, and the Minerva Foundation 
for support. A.T. acknowledges support from the 
Italian PRIN ``Fenomeni quantistici collettivi: 
dai sistemi fortemente correlati ai simulatori quantistici'' 
(PRIN 2010$\_$2010LLKJBX). M.B. acknowledges support from the EU grant SIQS.

\appendix

\section{Double-Weyl points as Berry monopoles} \label{app:monopole}

The effective Hamiltonian \eqref{Heff} describes the two 
internal bands in a neighborhood of one of the double-Weyl points 
and it allows to evaluate an approximated behavior of their associated 
Fermi arcs. The double-Weyl points are monopoles with charge two of the 
Berry flux associated to each of the two central bands. 
In order to show this we consider the particular case of $q=\pi/4$ and 
we evaluate the field $F=\vec{\nabla} \times \vec{\mathcal{A}}$ 
related to the Berry connection 
$\mathcal{A}_j=-i\bracket{\psi}{\partial_{\tilde{k}_j} \psi}$ calculated 
from the effective Hamiltonian \eqref{Heff}. We obtain:
\begin{align}
 F_x &=  \frac{\tilde k_x \left(\tilde k_x^2+\tilde k_y^2\right)}{\sqrt{2} \left(\tilde k_x^4-\tilde k_x^2 \tilde k_y^2+\tilde k_y^4+2 \tilde k_z^2\right)^{3/2}} \\
 F_y &=  \frac{\tilde k_y \left(\tilde k_x^2+\tilde k_y^2\right)}{\sqrt{2} \left(\tilde k_x^4-\tilde k_x^2 \tilde k_y^2+\tilde k_y^4+2 \tilde k_z^2\right)^{3/2}} \\
 F_z &=  -\frac{\sqrt{2} \, \tilde k_z \left(\tilde k_x^2+\tilde k_y^2\right)}{ \left(\tilde k_x^4-\tilde k_x^2 \tilde k_y^2+\tilde k_y^4+2 \tilde k_z^2\right)^{3/2}} \,.
\end{align}
By integrating the flux of $F$ for a closed surface around the Weyl point in 
$\vec{k}=(\pi/2,\pi/2,\pi/2)$ one obtains a flux $4\pi$, 
corresponding to a double charge.

\section{Overlap of the double-Weyl points and symmetries} \label{app:sym}

The overlap of double-Weyl points on the surface Brillouin zone along the 
canonical directions of the lattice is caused by  the 
symmetries of the Hamiltonian \eqref{Hpianiso}, including the momentum 
translations in \eqref{tras1} and \eqref{tras2}. Here we focus on the plane 
where the Weyl points lie, $k_z=\pi/2$, and list the symmetries of $H_{xy}$ 
that must be broken to avoid such overlaps.

First, there are anti-commuting reflection symmetries which map either 
$k_x\to-k_x$ or $k_y\to-k_y$:
\begin{align}
 H_{xy}\left(k_x,k_y \right)&= -\tau_z\sigma_y H_{xy}\left(-k_x,k_y \right) \tau_z\sigma_y \,, \label{rot1}\\ 
 H_{xy}\left(k_x,k_y\right)&= -\tau_z\sigma_x H_{xy}\left(k_x,-k_y \right) \tau_z\sigma_x \,. \label{rot2}
\end{align}
Then we must consider commuting reflection symmetries corresponding to the same reflections:
\begin{align}
  H_{xy}\left(k_x,k_y \right)&= \sigma_y H_{xy}\left(-k_x,k_y \right) \sigma_y \,,\label{ref1}\\ 
  H_{xy}\left(k_x,k_y \right)&= \sigma_x H_{xy}\left(k_x,-k_y \right) \sigma_x \,. \label{ref2}
\end{align}
Finally we must take into account commuting translational symmetries of 
the kind:
\begin{align}
 H_{xy}\left(k_x,k_y \right)&= \tau_y H_{xy}\left(k_x+\pi,k_y \right) \tau_y \,, \label{tras3}\\ 
 H_{xy}\left(k_x,k_y \right)&= \tau_x H_{xy}\left(k_x,k_y+\pi \right) \tau_x \,. \label{tras4}
 \end{align}

The effect of these sets of symmetries is that, given a Weyl 
point in a particular position, there will be two symmetric zero-energy 
points, with opposite monopole charge, which share respectively 
the same $k_x$ and $k_y$ component.

Therefore, analogously to the symmetries (\ref{tras1},\ref{tras2}), 
also the previous relations imply that, if we consider a boundary 
orthogonal to either the $\hat{x}$ or  the $\hat{y}$ direction, 
two Weyl points with opposite Berry monopole will overlap in the surface 
Brillouin zone and the related Fermi arcs will acquire a non-zero energy.

In order to avoid this, we have to introduce an additional term in 
the Hamiltonian, in such a way to break the symmetries 
(\ref{tras1},\ref{rot1},\ref{ref1},\ref{tras3}) to avoid overlaps 
in the $k_x$ direction, or to break 
(\ref{tras2},\ref{rot2},\ref{ref2},\ref{tras4}) in order 
to avoid the overlaps in the $k_y$ direction. 
As discussed in the main text, such an additional term necessarily 
breaks the $SU(2)$ gauge invariance and anti-commutes with $H_z$. 
Thus a simple Zeeman splitting is not sufficient. 
The term $H_{xx}$ in Eq. \eqref{stagzeeman} violates all the symmetries in 
the set (\ref{tras1},\ref{rot1},\ref{ref1},\ref{tras3}).

\end{document}